\documentstyle[12pt]{article}
\newcommand{\be}{\begin{equation}}
\newcommand{\ee}{\end{equation}}
\begin{document}

\title{Power Switching in Hybrid Coherent Couplers\\
\smallskip\ }
\author{W. D. Deering$^1$and M. I. Molina$^2$ \and \smallskip\  \\
{\it \ }$^1${\it \ Center for Nonlinear Science, Department of Physics}\\
{\it \ University of North Texas, Denton, TX 76203, USA}\\
\smallskip\ \\
{\it \ }$^2${\it Departamento de F\'\i sica, Facultad de Ciencias}\\
{\it \ Universidad de Chile}\\
{\it \ Casilla 653, Santiago, Chile}}
\date{}
\maketitle
\newpage\ 

\begin{center}
{\bf Abstract}
\end{center}

\smallskip\ 

We report on a theoretical and numerical investigation of the switching of
power in new hybrid models of nonlinear coherent couplers consisting of
optical slab waveguides with various orders of nonlinearity. The first model
consists of two guides with second-order instead of the usual third-order
susceptibilities as typified by the Jensen coupler. This second-order system
is shown to have a power self-trapping transition at a critical power
greater than the third-order susceptibility coupler. Next, we consider a
mixed coupler composed of a second-order guide coupled to a third-order
guide and show that, although it does not display a rigorous self-trapping
transition, for a particular choice of parameters it does show a fairly
abrupt trapping of power at a lower power than in the third-order coupler.
By coupling this mixed nonlinear pair to a third, purely linear guide, the
power trapping can be brought to even lower levels and in this way a
satisfactory switching profile can be achieved at less than one sixth the
input power needed in the Jensen coupler.

\newpage\ 

\section{ Introduction}

Interest in all-optical switching devices has led to the study and design of
several promising configurations of nonlinear couplers which display
intensity-triggered power switching. The basic nonlinear coherent coupler,
introduced by Jensen \cite{jensen}, consists of two similar waveguides made
of a material with third-order susceptibilities, embedded in a host with
purely linear susceptibility. When the guides are placed parallel to each
other and in close proximity over a given distance, the guide fields overlap
to some extent and power can be transferred between the two. When all the
power is initially launched into one of the guides, the nonlinear
susceptibility can give rise to self-trapping of power in the original
guide. The output power in the original guide, for a device length equal to
a coupling length, can be made to switch from essentially zero percent at
low power levels, to one hundred percent for input power levels exceeding a
characteristic threshold. In addition to the pioneering work by Jensen,
several other coupler configurations have been considered. It was found that
a three-in-a-line configuration of couplers displays a more abrupt switching
profile, at the expense however, of greater input power\cite{finlayson}. The
same tendency was reported for a linear array of many couplers\cite{schmidt}%
. In an effort to improve the switching profile, we introduced in a recent
work\cite{MDT-PD93} the Doubly Nonlinear Trimer (DNT) coupler consisting of
two nonlinear guides coupled to a third, linear guide. Such a system
displays the interesting phenomenon of power self-trapping {\em tunability}:
the critical input power level necessary for the onset of power
self-trapping can be tuned to low values, by adjusting the value of the
(linear) coupling between the nonlinear guides and the linear one.\cite
{MDT-PD93},\cite{MT-PRA92} In the optimal configuration, switching was
achieved at one-fourth the power needed to produce switching in the Jensen
coupler. The price to pay for this improved switching is the use of larger
device lengths, up to ten times that reported by Jensen\cite{MDT-PD93}.

In the present work, our interest is in learning if couplers having
waveguides with differing types of nonlinear susceptibilities would have
better switching characteristics than other standard models. We first
investigate a different nonlinear coupler composed of two identical guides
made of optical material lacking inversion symmetry and therefore having a
nonvanishing second-order susceptibility. We show that this new coupler
array possesses a power self-trapping transition and an associated sharp
power switching profile, albeit at a larger input power level than in
Jensen's and in our earlier DNT coupler. Then, after examining a number of
two-guide couplers of mixed compositions, with each guide having purely
linear (L), and second-order (SO) or the usual third-order (TO)
susceptibilities we found that for a particular choice of parameters, a
coupler composed of an SO guide and a TO guide displays a relatively sharp
power self-trapping profile at an input power level lower than previously
reported, if power is initially launched in the SO guide. Next, as in the
DNT case, the onset of self-trapping can be tuned to even lower power
levels, by perturbing the two-guide coupler by adding a purely linear
control guide and adjusting the strength of the interaction with this third
guide. The resulting three-guide coupler, dubbed SO-TO-L, resembles the DNT
configuration, with one of the third-order guides replaced by a second-order
guide; it displays a reasonably sharp switching profile and, as far as we
know, does so at the lowest input power reported so far.

\section{A new two-guide coupler}

Consider a linearly coupled system of two nonlinear guides, each having the
same second-order nonlinear susceptibility. In the single mode
approximation, the normalized mode amplitudes satisfy 
\begin{eqnarray}
i\frac{dC_1}{dz} &=&VC_2-\chi |C_1|C_1  \label{C1} \\
i\frac{dC_2}{dz} &=&VC_1-\chi |C_2|C_2,  \label{C2}
\end{eqnarray}
where $\chi =Q^{(2)}\sqrt{P}$ is the product of an integral $Q^{\left(
2\right) }$ containing the second-order nonlinear susceptibility\cite{jensen}
and the square root of the input power $P$. The linear coupling of the
guides is determined by the coefficient $V.$ With all the power initially
launched in guide $1$, the initial conditions are $C_1(0)=1$, $C_2(0)=0$. We
will now show that Eqns.(\ref{C1})-(\ref{C2}) predicts a {\em self-trapping}
of power in the original guide (guide $1$). First, it is convenient to
rewrite (\ref{C1}-\ref{C2}) as a set of four equations for the complex
quantities $\rho _{ij}\equiv C_iC_j^{*}$: 
\begin{eqnarray}
i\frac{d\rho _{11}}{dz} &=&-V(\rho _{12}-\rho _{21})  \label{r11} \\
i\frac{d\rho _{22}}{dz} &=&V(\rho _{12}-\rho _{21})  \label{r22} \\
i\frac{d\rho _{12}}{dz} &=&-V(\rho _{11}-\rho _{22})  \nonumber  \label{r12}
\\
&&+\chi (\sqrt{\rho _{22}}-\sqrt{\rho _{11}})\rho _{12}  \label{r12} \\
i\frac{d\rho _{21}}{dz} &=&V(\rho _{11}-\rho _{22})  \nonumber  \label{r21}
\\
&&-\chi (\sqrt{\rho _{22}}-\sqrt{\rho _{11}})\rho _{21}.  \label{r21}
\end{eqnarray}

We have two conserved quantities: the total power, normalized to unity: $%
\rho _{11}+\rho _{22}=1$ and the total ``energy'' $H=V(\rho _{12}+\rho
_{21})-(2/3)\chi (\rho _{11}^{3/2}+\rho _{22}^{3/2})=-(2/3)\chi $ leaving
only two independent unknowns, which precludes any chaotic dynamics for the
system. Making use of these conserved quantities we find, after some tedious
algebra, the following first-order equation for $\rho _{11}\equiv \rho $: 
\begin{equation}
{\frac 1{{2}}}({\frac{d\rho }{{dz}}})^2+U(\rho )=0  \label{energy}
\end{equation}
with 
\begin{eqnarray}
U(\rho ) &=&-2\rho (1-\rho )+{\frac 1{{2}}}\left( {\frac{2\chi }{{3V}}}%
\right) ^2  \nonumber  \label{U} \\
&&\ -{\frac 2{{3}}}\left( {\frac \chi {{V}}}\right) ^2\sqrt{1-\rho }\left( -{%
\frac 2{{3}}}\rho ^{3/2}(1-\rho )\right)  \nonumber \\
&&\ -{\frac 1{{3}}}(\rho ^3+(1-\rho )^3)  \nonumber  \label{eq:5} \\
&&\ +{\frac 2{{3}}}(\rho ^{3/2}+(1-\rho )^{3/2}).  \label{U}
\end{eqnarray}
Equation (\ref{energy}) describes a classical particle of unit mass, moving
under the influence of an external potential $U(\rho )$, with initial
condition $\rho (0)=1$. Fig.1 shows the effective potential $U(\rho )$ for
several different values of $\chi /V$. For small nonlinearity values, the
effective potential is concave and conservation of energy allows complete
oscillations of the ``particle''; that is, power is transferred between the
two guides. As nonlinearity (input power) is increased, the potential
develops a local maximum whose height increases with increasing
nonlinearity. The condition for self-trapping of power in the original guide
translates here into the condition for the potential $U(\rho )$ to develop a
double root at $\rho =\rho ^{*}$ for some critical value of $\chi /V$, i.e., 
$U(\rho ^{*})=0$ and $(dU/d\rho )_{\rho ^{*}}=0$. Close examination of Eq.(%
\ref{U}) and Fig.1 reveals $U(\rho )$ to be {\em even} around $\rho =1/2$
and that $\rho ^{*}=1/2$. From that, the critical value of the nonlinearity
is obtained in closed form as 
\begin{equation}
\left( {\frac \chi {{V}}}\right) _c={\left( \frac 3{{\sqrt{2}}}\right) }%
\sqrt{3+2\sqrt{2}}\approx 5.121.  \label{eq:6}
\end{equation}
This value is greater than the critical values for Jensen's coupler ($=4$)
and for the array of three nonlinear (third-order) couplers$^2$ ($\approx
4.5 $). Figure 2 shows the average transmittance of the guide, defined as 
\begin{equation}
<P>\equiv \lim_{L\rightarrow \infty }(1/L)\int_0^L\rho (z)dz.  \label{<P>}
\end{equation}
Clearly, we see that for $(\chi /V)<(\chi /V)_c$, power is equally
distributed between the two guides. At $(\chi /V)=(\chi /V)_c$, an abrupt
transition takes place and power begins to self-trap in the original guide.
Onset of self-trapping is a precursor for the appearance of a sharp
switching profile in the transmittance of the guide. The transmittance,
defined as $|C_1(L_c)|^2,$ is the quantity of basic interest for optics. The
length $L_c$ is usually chosen as the shortest length for which $|C_1(z)|^2$
is zero, or very nearly so, in the absence of nonlinearity ($\chi =0$). In
the case of the two waveguide system, $L_c=\pi /(2V)$. The abrupt increase
in transmittance caused by an increment of the nonlinearity parameter (input
power) can be used as a power triggered switch\cite{jensen}.

Figure 3 shows the transmittance characteristics of our two-guide
second-order (SO) coupler, and compares it with Jensen's third-order (TO)
nonlinear coupler which is also shown in the figure, along with the TO
nonlinear coupler with three guides\cite{finlayson}. We note the SO
nonlinear coupler array does not have a competitive switching profile
compared to Jensen's and the three-coupler array.

\section{A New Hybrid Configuration}

After considering the above nonlinear coupler, having second-order
susceptibility, we next examined a variety of mixed two-guide couplers in
which each guide was either a purely linear one, a SO or a TO guide. The
objective was to find other two-guide couplers that displayed power
self-trapping for the initial condition where all the initial power is put
into one guide. We found that, in most cases there is no self-trapping
transition at all but a continuous power trapping. For a given mixed
two-guide coupler, the trapping profile depends in a sensitive way on the
order of the nonlinear susceptibility of the guide initially receiving all
power. To illustrate this point, we now describe the most interesting case
we found: The SO-TO guide system, where guide $1$ possesses a second-order
nonlinear susceptibility integral\cite{jensen} $Q_1^{(2)}$ and guide $2$
possesses the usual third-order susceptibility integral\cite{jensen} $%
Q_2^{(3)}$. The equations for the mode amplitudes are 
\begin{eqnarray}
i\frac{dC_1}{dz} &=&VC_2-\chi _1|C_1|C_1  \label{pepe1} \\
i\frac{dC_2}{dz} &=&VC_1-\chi _2|C_2|^2C_2,  \label{pepe2}
\end{eqnarray}
where $\chi _1=Q_1^{(2)}\sqrt{P}$ and $\chi _2=Q_2^{(3)}P$. When all initial
input power goes into the TO guide (\#2), the initial condition for the
system, Eqns. (\ref{pepe1})-(\ref{pepe2}), is $C_1(0)=0$, $C_2(0)=1$. A
numerical investigation of $<P>$ reveals a ``delayed'' self-trapping
transition at $\chi _1=\chi _2=\chi _c\approx 6.3\ V$ (Fig.4). This value is
much greater than Jensen's and is, therefore, not useful for our purposes.
On the other hand, when all input power is put initially into the SO guide
(\#1), we have the initial condition $C_1(0)=1$, $C_2(0)=0$. In this case, a
numerical search reveals that this system does not show a self-trapping
transition: the effective potential $U(\rho ,\chi _1,\chi _2)$ does not
develop a double root for any combination of $\chi _1,\chi _2$. However, for
the special case $\chi _1=\chi _2\equiv \chi $, we found a relatively sharp
power self-trapping profile occurring at $\chi \approx 3.0\ V$ (Fig.4); {\em %
i.e.}, a smaller power than Jensen's critical value for self-trapping. We
then proceeded to ``tune'' the trapping profile to even lower power levels,
by allowing the SO-TO coupler to interact linearly with a third (control)
guide possessing only linear susceptibility. The enlarged set of equations
for the mode amplitudes in this SO-TO-L coupler now reads 
\begin{eqnarray}
i\frac{dC_1}{dz} &=&VC_2+WC_3-\chi |C_1|C_1  \label{SOTOL1} \\
i\frac{dC_2}{dz} &=&VC_1+WC_3-\chi |C_2|^2C_2  \label{SOTOL2} \\
i\frac{dC_3}{dz} &=&W(C_1+C_2),  \label{SOTOL3}
\end{eqnarray}
with initial conditions $C_1(0)=1$, $C_2(0)=C_3(0)=0$. It is assumed here
that the guides have the same {\em linear} susceptibility, to minimize
possible phase mismatch effects. After examining $<P>$ as a function of $%
\chi $ for different $W$ values, we found that $W\approx 1.1V$ brings the
onset of self-trapping down to a power level $\chi \approx 0.4\ V$. Note
that this optimal $W$ value is the same as found for the DNT coupler\cite
{MDT-PD93}. Now, to evaluate the transmittance of this SO-TO-L array, we
need to calculate the coupling length $L_c(W)$. This is obtained from Eqns. (%
\ref{SOTOL1})-(\ref{SOTOL3}) as the position $z$ at which $|C_1(z)|^2\approx
0$, for $\chi =0$. In this limit the system of equations can be solved in
closed form\cite{MDT-PD93} and yields for $|C_1(z)|^2$: 
\begin{eqnarray}
|C_1(z)|^2 &=&A\cos \left[ \left( \frac{3V-\sqrt{V^2+8W^2}}{{2}}\right)
z\right]   \nonumber \\
&&\ \ \ \ \ \ \ +B\cos \left[ \left( \frac{3V+\sqrt{V^2+8W^2}}{{2}}\right)
z\right]   \nonumber \\
&&\ \ \ \ \ \ \ +{\em C}\cos [\sqrt{V^2+8W^2}z]+{\em D,}  \label{c1sqr}
\end{eqnarray}
where 
\[
A=\left( \sqrt{V^2+8W^2}-V\right) /\left( 4\sqrt{V^2+8W^2}\right) 
\]
\[
B=\left( \sqrt{V^2+8W^2}+V\right) /\left( 4\sqrt{V^2+8W^2}\right) 
\]
\[
C=W^2/\left( V^2+8W^2\right) 
\]
\[
D=\left( V^2+4W^2\right) /\left[ 4\left( V^2+8W^2\right) \right] +1/4.
\]
For $W=1.1\ V,$ Eqn.(\ref{c1sqr}) gives $L_c\approx 21/V$, the same value as
for the DNT coupler. Figure 5 shows the transmittance of the SO-TO-L system
as a function of input power, for the optimal linear coupling value $W=1.1V$%
. For comparison we also show the transmittance for the DNT coupler.
Jensen's device switches at about $\chi =4V$ and the side-by-side
three-nonlinear guide coupler of ref. 2 switches at about $\chi \sim 4.5\ V,$
but because of the scale of the figure, neither of these transitions is
shown. We note that the new coupler configuration SO-TO-L is capable of
achieving over $99\%$ power switching for input power levels below $\chi
\sim 0.65\ V$ which is a $48\%$ reduction in input power needed compared to
the DNT device.

\section{Discussion}

In order for the above results to be meaningful, it must be true that $\chi
_2$ and $\chi _3$ can be at least approximately equal for some materials.
These coefficients involve the usual susceptibilities $\chi ^{\left(
j\right) }$ defined here to give the electric polarization $P_{E}$ in
the form 
\[
P_{Ei}=\epsilon _0\left[ \chi _{ij}^{\left( 1\right) }E_j+\chi
_{ijk}^{\left( 2\right) }E_jE_k+\chi _{ijkm}^{\left( 3\right)
}E_jE_kE_m+\cdots \right] . 
\]
To find the ratio $\chi _2/\chi _3=$ $Q_2/\left( Q_3\sqrt{P}\right) ,$ we
use the definitions from ref. 1 of the integrals $Q_2$ and $Q_3,$ inserting
the exact expressions for mode fields and susceptibilities. Rather than
going through those calculations, we make the simplifying assumptions that
the $\chi ^{\left( j\right) }$ are constant across each guide and that the
mode field is also constant (approximately true for the $TE_0$ mode) across
the guide; then the integrals are easily done and we get 
\[
\chi _2/\chi _3\simeq \frac{\chi ^{\left( 2\right) }}{\chi ^{\left( 3\right)
}\left| E\right| \sqrt{P}}, 
\]
where $P$ is the input power and $\left| E\right| $ is the amplitude of a
slab waveguide mode field, normalized to one watt/meter. Then the ratio $%
\chi _2/\chi _3$ can be on the order of unity within the range of known
values of the susceptibilities\cite{boyd} and power in the range 0.01 - 1 kw.

As mentioned previously, the critical length $L_c$ for the SO coupler is the
same as for the Jensen coupler, but the SO\ device switches less abruptly
and at higher power than Jensen's. The SO-TO coupler shows final-state
asymmetry depending on which guide receives input power. If power enters the
TO leg, a self-trapping transition occurs at more than 1.5 times the Jensen
level, $P_J.$ If the SO\ leg receives the power, a relatively sharp
self-trapping sets in at about 25\% below $P_J.$

A greatly lowered power switching level is shown by SO-TO-L, but its $L_c$
is an order of magnitude larger than the Jensen $L_c.$ Typical values for $%
L_c$ are about a millimeter\cite{yeh} for weakly coupled devices ( $i.e.,$
the separations between waveguides are large enough that coupled-mode theory
can be used) and less for stronger coupling. Then $L_c$ for SO-TO-L is on
the order of a centimeter or less.

The linear interaction coefficients $V$ and $W$ are overlap integrals,
across one waveguide, of the product of the electric mode field of that
guide and the mode field of a second guide. Therefore, $V$ and $W$ are
functions of the separation of the waveguides and in principle, it is
possible to alter one without changing the other; that is, the system can be
tuned to achieve minimum power switching level, by changing the distances
between the linear guide and the other two, nonlinear guides.

\section{Conclusions}

Our primary interest was the investigation of switching characteristics of
model nonlinear couplers having mixtures of waveguides, not necessarily with
the same orders of nonlinear susceptibilities. Earlier work on the DNT
system suggested tunability might also be used in a hybrid coupler to
decrease switching power levels. It appears possible to meet the condition $%
\chi _2\cong $ $\chi _3,$ as far as known values of these quantities are
concerned. Whether specific materials can be found that meet this condition
and are also compatible with one another in a device, is another matter and
one we have not addressed in this paper .

Switching characteristics of SO is inferior to the TO system. For SO-TO, the
asymmetry of final states with respect to input guide may be the only aspect
of its performance that could be of interest.

The most interesting coupler was the SO-TO-L, formed by adding a linear
guide to SO-TO and tuning for minimum power by adjusting the relative
positions of the guides. The transition power level drops to less than
one-sixth of $P_J.$ Although a disadvantage of this coupler is a critical
length that is longer than for the Jensen coupler by an order of magnitude,
that may be tolerable in some applications.

Of course, there are various other configurations involving arrays of these
couplers but those were not investigated.

One of the authors (M. I. M.) acknowledges support from FONDECYT grant
1950655.

\newpage

\centerline{FIGURE CAPTIONS}

\vspace{1.0 cm}

\noindent {\bf Figure 1:}\ SO coupler system:\ Effective classical potential 
$U(\rho )$ for a particle of unit mass and ``coordinate'' $\rho $ which, at $%
z=0$ starts from $\rho =1$. The condition for the onset of a power
self-trapping transition is the appearance of a double root at some value of
nonlinearity (input power). For our SO coupler, $(\chi /V)_c=5.12$.\\

\noindent {\bf Figure 2:}\ \ Space-averaged transmittance $<P>$ for the SO
coupler (solid line). At $(\chi /V)=5.12$ there is a power self-trapping
transition and power begins to selftrap in the original guide. For
comparison we also show $<P>$ for Jensen's coupler (dashed line).\\

\noindent {\bf Figure 3:}\ \ Transmittance $|C_1(L_c)|^2$ versus the power
parameter $\chi /V$ for the SO nonlinear coupler (solid line, $L_c=\pi /2\ V$%
), Jensen's coupler (dotted line, $L_c=\pi /2\ V$) and the three-in-a-line
TO nonlinear coupler (dashed line, $L_c=\pi /\sqrt{2}\ V$).\\

\noindent {\bf Figure 4:}\ \ Space-averaged transmittance for the SO-TO
coupler. When all the initial input power goes to the TO guide, we have a
``delayed'' self-trapping transition around $\chi /V\approx 6.3$ (dashed
line). If the initial power is put into the SO guide, there is no
self-trapping transition, but at $(\chi /V)\approx 3.0$ power begins to
self-trap in the original guide in a reasonably sharp manner (solid line). \\

\noindent {\bf Figure 5:}\ \ Transmittance $|C_1(L_c)|^2$ versus power
parameter $\chi /V$ for the SO-TO-L coupler (solid line) and the DNT coupler
(dashed line). In both cases $L_c\approx 21/V$.

%
\begin{figure}[p]
\begin{center}
\leavevmode
\hbox{
\includegraphics{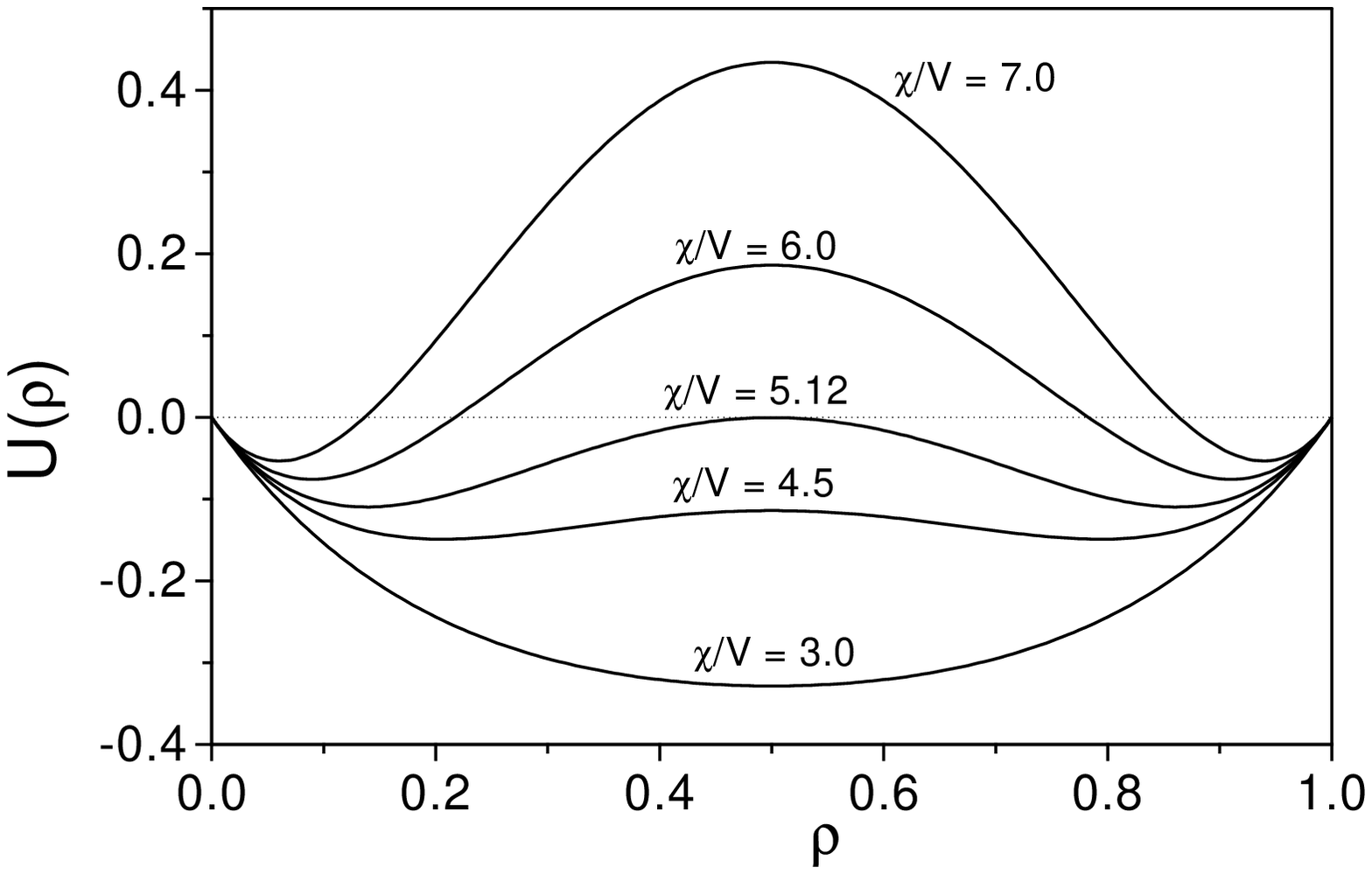}  }	
\end{center}
\vspace{2.3 in}
\label{figure1}
\vspace{4.0cm}
\centerline{\small{Fig.1}}
\end{figure}
%
\begin{figure}[p]
\begin{center}
\leavevmode
\hbox{
\includegraphics{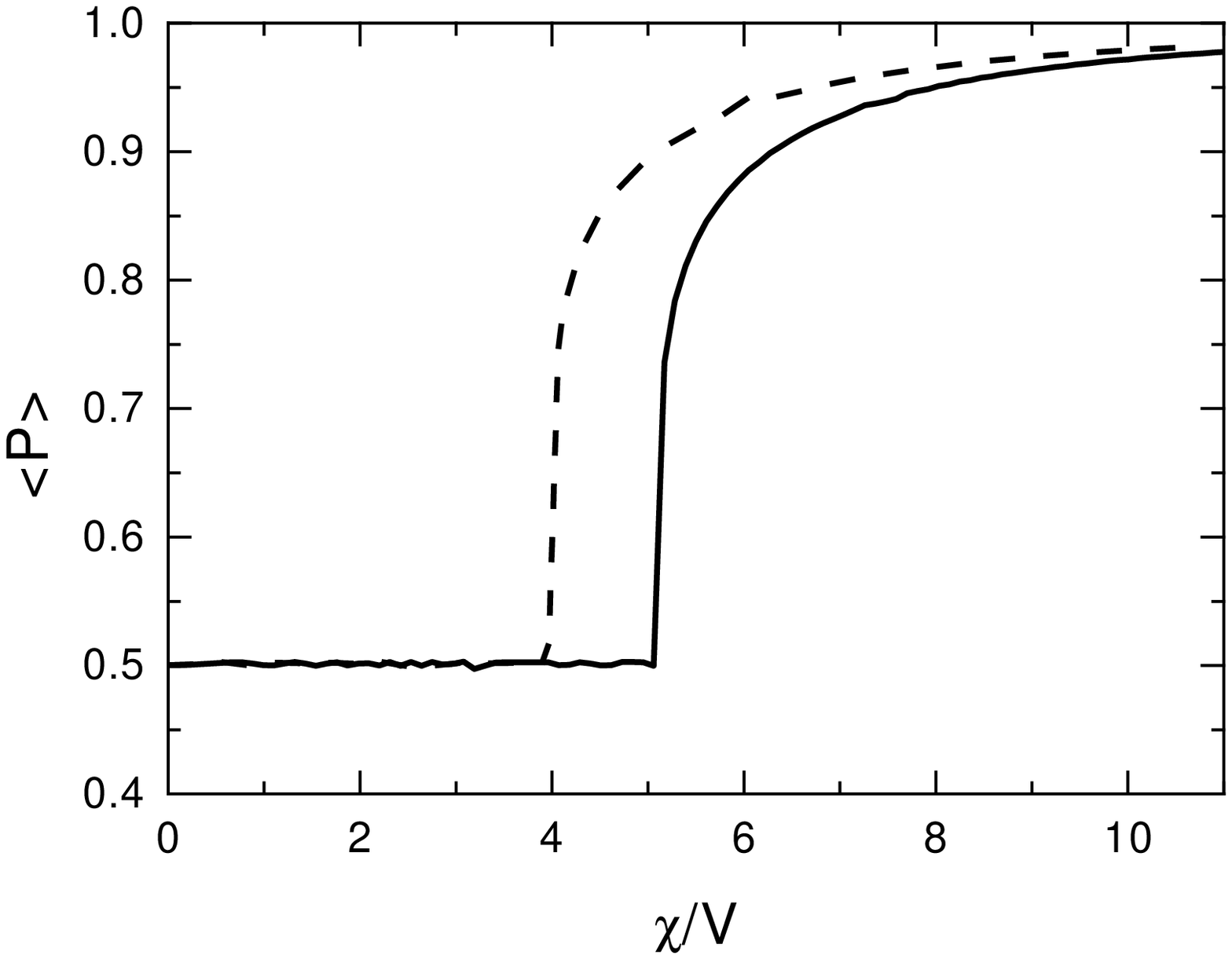}  }	
\end{center}
\vspace{2.3 in}
\label{figure2}
\vspace{6.0cm}
\centerline{\small{Fig.2}}
\end{figure}
%
\begin{figure}[p]
\begin{center}
\leavevmode
\hbox{
\includegraphics{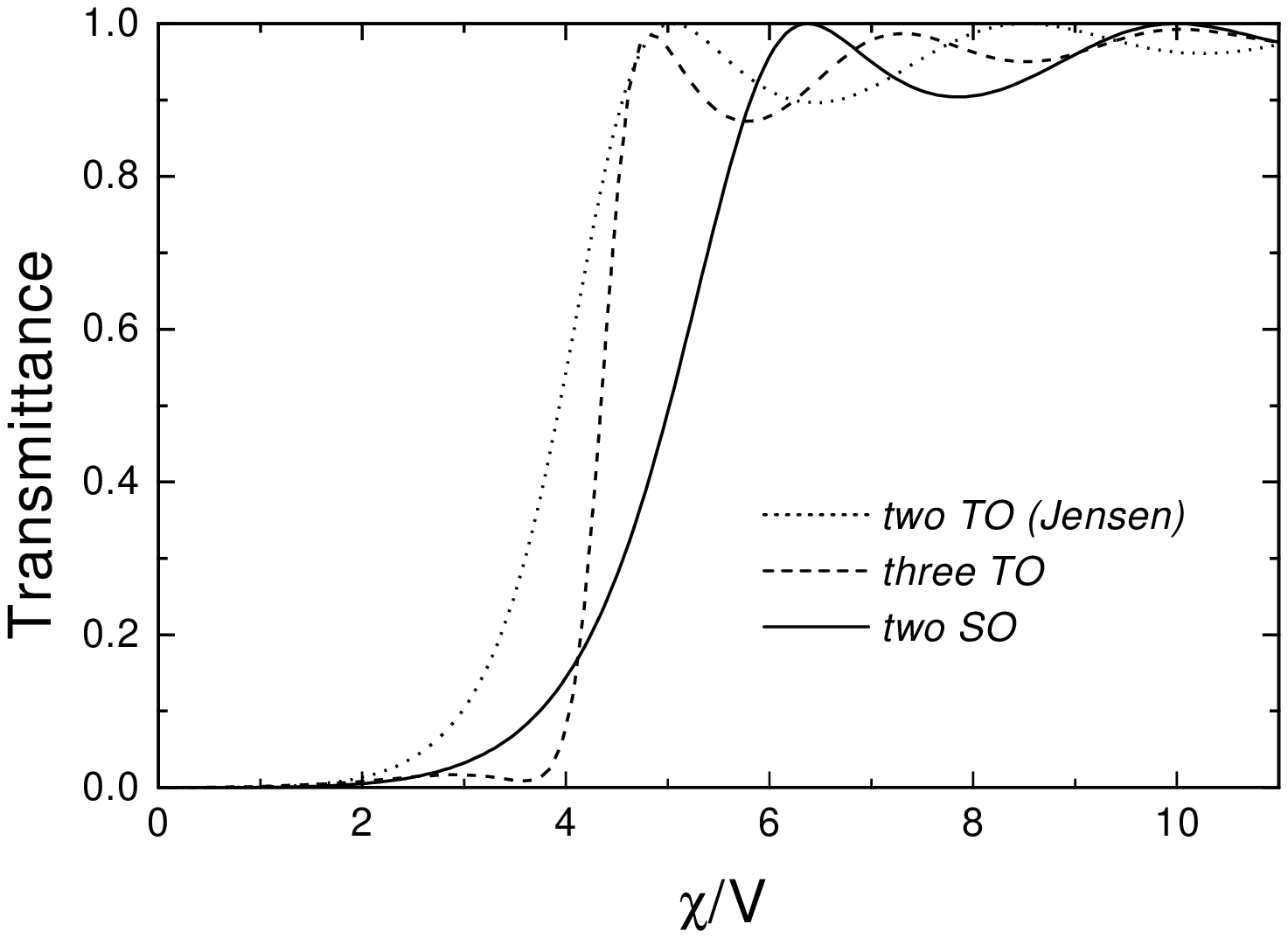}  }	
\end{center}
\vspace{2.3 in}
\label{figure3}
\vspace{6.0cm}
\centerline{\small{Fig.3}}
\end{figure}
%
\begin{figure}[p]
\begin{center}
\leavevmode
\hbox{
\includegraphics{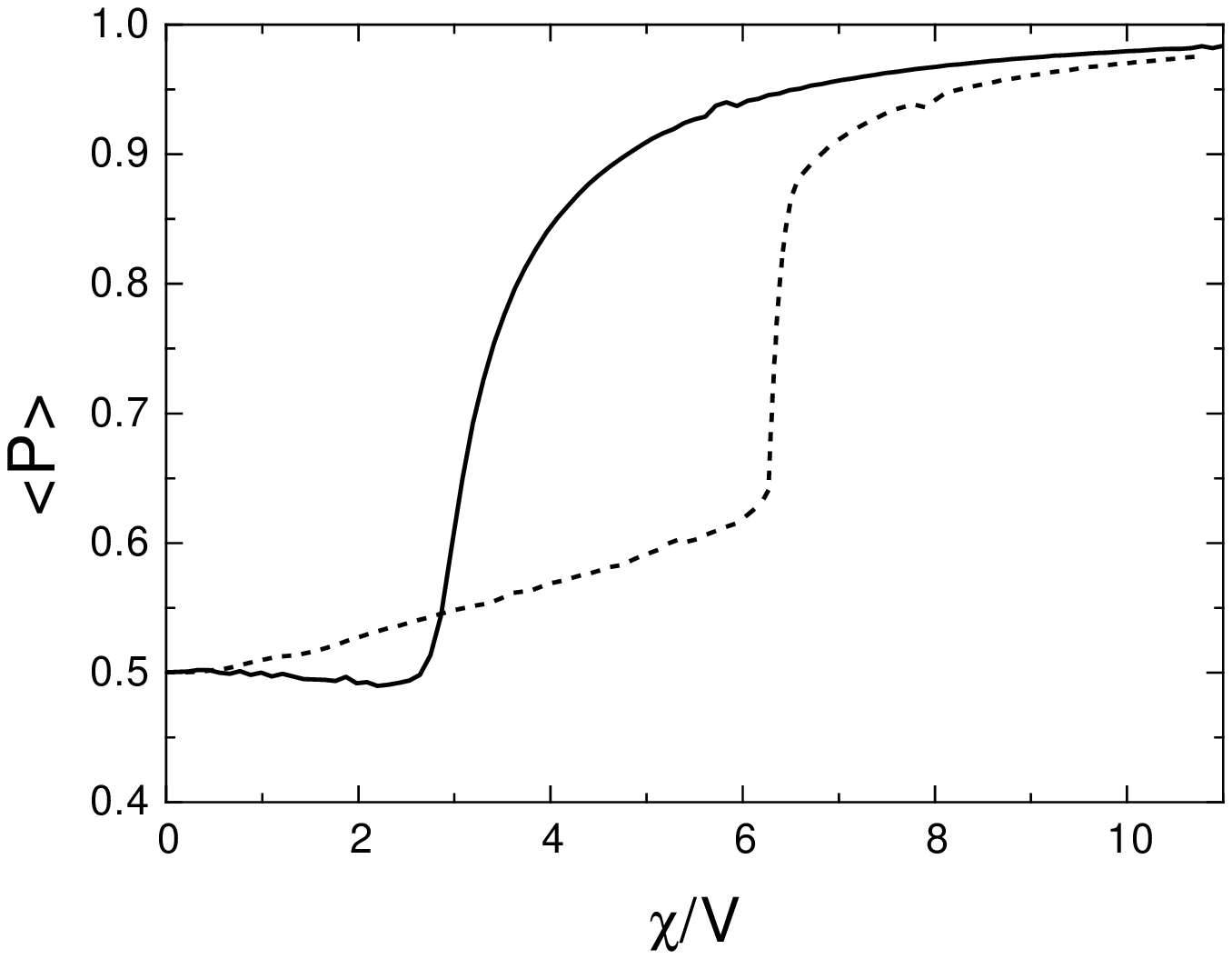}  }	
\end{center}
\vspace{2.3 in}
\label{figure4}
\vspace{6.0cm}
\centerline{\small{Fig.4}}
\end{figure}
%
\begin{figure}[p]
\begin{center}
\leavevmode
\hbox{
\includegraphics{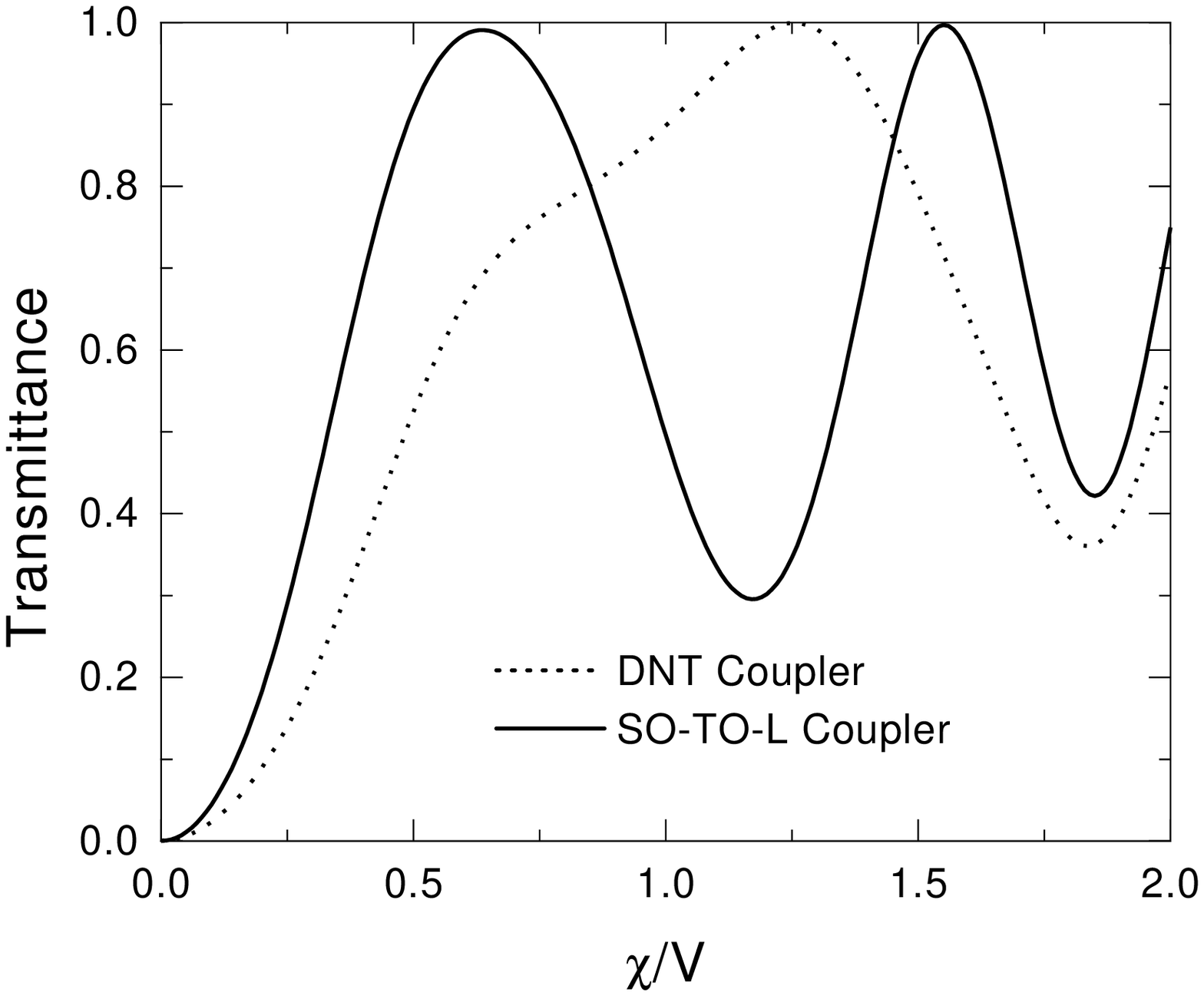}  }	
\end{center}
\vspace{2.3 in}
\label{figure5}
\vspace{7.0cm}
\centerline{\small{Fig.5}}
\end{figure}

\end{document}